\documentclass[useAMS,usenatbib,usegraphicx]{mn2e}
\usepackage{graphicx}

\renewcommand\theequation{\arabic{equation}}
\newcommand{\bea}{\begin{eqnarray}}
\newcommand{\eea}{\end{eqnarray}}

\title[]{Testing the local-void alternative to dark energy using galaxy pairs}
\author[F. Y. Wang and Z. G. Dai] {F. Y. Wang \thanks{fayinwang@nju.edu.cn} and Z. G. Dai\thanks{dzg@nju.edu.cn}\\
School of Astronomy and Space Science, Nanjing University, Nanjing
210093, China\\
Key laboratory of Modern Astronomy and Astrophysics (Nanjing
University), Ministry of Education, Nanjing 210093, China}
\begin{document}

\maketitle


\begin{abstract}
The possibility that we live in a special place in the universe,
close to the center of a large, radially inhomogeneous void, has
attracted attention recently as an alternative to dark energy or
modified gravity to explain the accelerating universe. We show that
the distribution of orientations of galaxy pairs can be used to test
the Copernican principle that we are not in a central or special
region of Universe. The popular void models can not fit both the
latest type Ia supernova, cosmic microwave background data and the
distribution of orientations of galaxy pairs simultaneously. Our
results rule out the void models at the $4\sigma$ confidence level
as the origin of cosmic acceleration and favor the Copernican
principle.
\end{abstract}


\section{Introduction}
The standard model of cosmology based on the cosmological principle
(homogeneity, isotropy, validity of General Relativity) which
contains about 23\% dark matter, 4\% ordinary matter and 73\% dark
energy driving the acceleration of a flat universe has been
established. Many astronomical observations support this standard
picture, including type Ia supernovae (SNe Ia) (Riess et al. 1998;
Perlmutter et al. 1999), cosmic microwave background (CMB) (Komatsu
et al. 2011; Sherwin et al. 2011), baryon acoustic oscillations
(BAO) (Eisenstein et al. 2005) and gamma-ray bursts (Dai et al.
2004; Wang et al. 2007; 2011).

In the meanwhile, inhomogeneous Lema\^{\i}tre-Tolman-Bondi (LTB)
(Lema\^{\i}tre 1933; Tolman 1934; Bondi 1947) universe could also
induce an apparent dimming of the light of distant supernovae. The
idea is to drop the dark energy and the Copernican principle, and
instead suppose that we are near the center of a large, nonlinearly
underdense, nearly spherical void surrounded by a flat, matter
dominated Einstein-de Sitter (EdS) spacetime. Because the observer
must be at the center of void, so the LTB models violate the
Copernican principle. Because of the observed isotropy of the CMB,
the observer must be located very close to the center of the void
(Alnes \&Amarzguioui 2006). It was demonstrated LTB models can fit
the SNe Ia data, as well as the BAO data and the CMB data
(Garcia-Bellido \& Haugboelle 2008). Some tests have actually been
proposed: the Goodman-Caldwell-Stebbins test, which looks at the CMB
inside our past lightcone (Goodman 1995; Caldwell \& Stebbins 2008),
the curvature test, which is based on the tight relation between
curvature and expansion history in a Friedmann spacetime (Clarkson
et al. 2008), and the radial and transverse BAO scale (Zibin et al.
2008; Garcia-Bellido \& Haugboelle 2009). However, based on these
tests, void models have not yet been ruled out (Clifton et al. 2008;
Uzan et al. 2008; Biswas et al. 2010; Wang \& Zhang 2012; Nadathur
\& Sarkar 2011). Zhang \& Stebbins (2011) have excluded the Hubble
bubble model as the possibility of cosmic acceleration using the the
Compton-y distortion. Zibin \& Moss (2011) also concluded that a
very large class of void models was ruled out using this method.
Here we propose a powerful tool, the orientations of galaxy pairs to
test the Copernican principle.

The Alcock-Paczynski (AP) test is a purely geometric test of the
expansion of the Universe (Alcock \& Paczyski 1979). Marinoni \&
Buzzi (2010) implemented the AP test with the distribution of
orientations of galaxy pairs in orbit around each other in binary
systems. The principle of this method is that the orientations is
thought to be completely random, with all orientations being equally
likely if measured assuming a cosmology that matches the true
underlying cosmology of the Universe in a
Friedmann-Lema\^{\i}tre-Robertson-Walker (FLRW) universe after the
effect of peculiar motion is excluded.

In this paper, we implement the Alcock-Paczynski test with pairs of
galaxies to test the Copernican principle. The void models cannot
both fit SNe Ia plus CMB data and orientations of galaxy pairs. Our
results exclude the possibility of the void models as the source of
cosmic accelerating expansion and favor the Copernican principle.

\section{The Void model}
We model the void as an isotropic, radially inhomogeneous universe
described by the LTB metric,
\begin{equation}
ds^2 = -c^2dt^2 + \frac{A^{\prime2}(r, t)}{1 + k(r)} dr^2
 + A^2(r, t)d\Omega^2,
\end{equation}
where a prime denotes the partial derivative with respect to the
coordinate distance $r$, and the curvature $k(r)$ is a free function
representing the local curvature. The transverse expansion rate is
defined as $H_{\perp} \equiv \dot{A}(r,t)/A(r,t)$ and the radial
expansion rate is defined as $H_{\parallel}\equiv \dot{A}^\prime(r,
t)/A^\prime(r, t)$, where an overdot denotes the partial derivative
with respect to $t$.

The Friedmann equation in LTB metric is $H_\perp^2 = F(r)/A^3(r, t)
+ c^2k(r)/A^2(r, t)$, where $F(r)>0$ is a free function which
determines the local energy density. The dimensionless density
parameters can be determined as $\Omega_M(r)$ and $\Omega_K(r)$ by
$F(r)=H_0^2(r)\Omega_M(r)A_0^3(r)$ and $c^2k(r) =
H_0^2(r)\Omega_K(r)A_0^2(r)$, where $H_0(r)$ and $A_0(r)$ are the
values of $H_\perp(r,t)$ and $A(r,t)$ respectively at the present
time $t=t_0$. So we can rewrite Friedmann equation in LTB metric as
$H_\perp^2(r,t)=H_0^2(r)[\Omega_M(r)(A_0/A)^3 +
\Omega_K(r)(A_0/A)^2]$. This equation can be integrated from the
time of the Big Bang, $t_\mathrm{B}=t_\mathrm{B}(r)$, to yield the
age of the universe at any given $(r, t)$,
\begin{equation}
t_0 - t_\mathrm{B}(r) = \frac{1}{H_0(r)} \int_0^{A/A_0}
 \frac{\mathrm{d}x}{\sqrt{\Omega_M(r) x^{-1} + \Omega_K(r)}}.
\end{equation}
The function, $A_0(r)$, corresponds to a gauge mode and we choose to
set $A_0(r)=r$. As stressed by Silk (1977) and Zibin (2008), it is
crucial to consider only voids with vanishing decaying mode, so we
set $t_\mathrm{B}(r)=0$ everywhere. Although Biswas et al (2010)
have shown that the void models were in better agreement with
observations if the void has been generated sometime in the early
universe. The null radial geodesics described by
\begin{eqnarray}
\label{geodesics}
\frac{dt}{dz}=-\frac{1}{\left(1+z\right)H_\parallel(z)}
,\label{geodesics:t}
\frac{dr}{dz}=\frac{c\sqrt{1+k\left(r\right)}}{\left(1+z\right)A^\prime(z)H_\parallel(z)},
\label{geodesics:r}
\end{eqnarray}
where $H_\parallel(z)=H_\parallel\left(r(z),t(z)\right)$. The
angular diameter distance and luminosity distance are given by
\begin{equation}
\label{d_A} d_A(z)=A\left(r(z),t(z)\right) ,
d_\mathrm{L}(z)=\left(1+z\right)^2A\left(r(z),t(z)\right) .
\end{equation}

We will adopt the two parameterizations of the void profile
$\Omega_M(r)$. The first one is the constrained GBH model
(Garcia-Bellido \& Haugboelle 2008)
\begin{equation} \label{omegam} \Omega_M(r)=\Omega_{M,
out}+(\Omega_{M, in}-\Omega_{M,out}) \, \frac{1-\tanh ( r-r_0 /
2\Delta r )}{1+\tanh (r_0/ 2\Delta r )} \,,
\end{equation}
where the parameters $r_0$ and $\Delta r$ characterize size and
steepness of the density profile respectively. We wish to look only
at voids that are asymptotically EdS, so we set
$\Omega_{M,\mathrm{out}}=1$. We also set $\Delta r=0.35 r_0$,
because this value can well fit the SNe Ia data (Garcia-Bellido \&
Haugboelle 2008; Marra \& Paakkonen 2010). This density shape of GBH
model can also explain other observations, such as CMB and BAO. The
second one is a simple Gaussian form,
\begin{equation}
\label{Om(r)}
\Omega_M(r)=\Omega_{M,\mathrm{out}}+\left(\Omega_{M,\mathrm{in}}-\Omega_{M,\mathrm{out}}\right)\exp(-r^2/r_0^2),
\end{equation}
where $\Omega_{M,\mathrm{in}}$ and $\Omega_{M,\mathrm{out}}$ are the
matter density parameters at the observer's position and in the FLRW
background outside the void, and $r_0$ characterizes the size of the
void. It has been shown that this void profile can fit the
observations of SNe Ia, CMB and BAO (Nadathur \& Sarkar 2011).

\section{Constraint from SNe Ia, CMB and pairs of galaxies}

We use the recent Union2 SNe Compilation (Amanullah et al. 2010),
which consists 557 SNe Ia in the redshift range $z=0.015-1.4$. With
$d_{L}$ in units of megaparsecs, the predicted distance modulus is
$\mu(z)=5\log d_{L}(z)+25$. The likelihood analysis is based on the
$\chi^2$ function:
\begin{equation}
\chi'^2_{\rm SNe}=\sum_{i=1}^{557} \frac{[\mu(z_i)-\mu_{\rm
obs}(z_i)+\mu]^2}{\sigma_i^2}.
\end{equation}
The parameter $\mu$ is an unknown offset. We marginalize the
likelihood $\exp (-\chi'^2_{\rm SNe}/2)$ over $\mu$, leading to a
new marginalized $\chi^2$ function:
\begin{equation}
\chi_{\rm SNe}^2 = S_2-\frac{S_1^2}{S_0} \,,
\end{equation}
where $S_n=\sum_i[\mu(z_i)-\mu_{\rm obs}(z_i)]^n/\sigma_i^2$.

\begin{figure}
\includegraphics[width=0.4\textwidth]{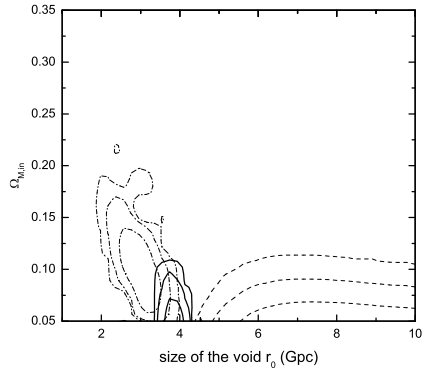}
\caption{\label{Fig1} The $1\sigma$, $2\sigma$ and $3\sigma$
contours in the parameter space $\Omega_{M,\rm in}-r_0$ for the
constrained GBH model. The dash-dot contours represent constraint
from SNe Ia+CMB, the dashed contours from AAP, and the solid
contours from SNe Ia+CMB+AAP. The best fit parameters are $r_0=3.0$
Gpc and $\Omega_{\rm M,in}=0.10$ for SNe Ia+CMB. But in order to fit
the AAP, much more larger and underdense void are needed. This void
model can not fit both the SNe Ia plus CMB data and the orientations
of galaxy pairs.}
\end{figure}

We also use positions and amplitudes of peaks and troughs in the CMB
spectrum to test the LTB models. The location of peaks and troughs
can be calculated as Hu et al. (2001): $l_m=(m-\phi_m) \, l_A \,$,
where $l_A=\pi \, \frac{d_{A}(z^{*}) (1+z^{*})}{r_s^*}$, where
$d_{A}(z^{*})$ is the angular diameter distance with the sound
horizon of $r_s^*$ at the recombination redshift of $z^{*}$. We use
the method of Marra \& Paakkonen (2010) to calculated these values.
We consider the position of the first, second, third peak and of the
first trough. We compute the corresponding phases $\phi_{1}$,
$\phi_{1.5}$, $\phi_{2}$ and $\phi_{3}$ using the accurate
analytical fits of Doran \& Lilley (2002). The relative heights of
second and third peak relative to the first one, $H_{2}$ and $H_{3}$
are also considered, for which we can use the fits of Hu et al.
(2001). So the $\chi^2_{\rm CMB}$ is (Marra \& Paakkonen 2010)
\begin{eqnarray} \label{chi2cmb}
\chi^2_{\rm CMB}=\sum_{1,1.5,2,3}
 \frac{(l_m-l_{m, W7})^2}{\sigma_{l_m}^2} + \sum_{2,3}\frac{(H_j-H_{j, W7})^2}{\sigma_{H_j}^2},
\end{eqnarray}
where the W7 represents the best-fit WMAP7 spectrum (Jarosik et al.
2011).

\begin{figure}
\includegraphics[width=0.4\textwidth]{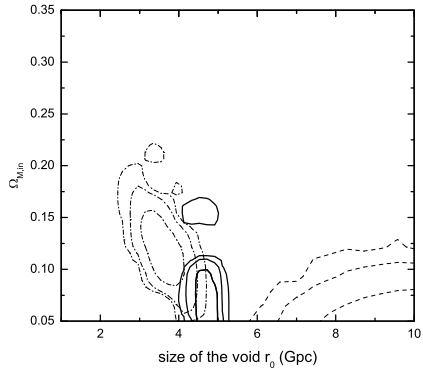}
\caption{\label{Fig2} Same as Fig.2, but for the gaussian LTB model.
The best fit parameters are $r_0=3.6$ Gpc and $\Omega_{\rm
M,in}=0.12$ from SNe Ia+CMB. The AAP favors a much more larger and
underdense void.}
\end{figure}

Pairs of galaxies should be distributed with random orientations if
the fundamental assumptions of homogeneity and isotropy are correct.
But two factors affect this simple cosmology test. First, peculiar
velocities displace the position of a galaxy along the line of sight
from its true position. Marinoni and Buzzi modelled the peculiar
velocity distortion as a Doppler shift where the observed line of
sight separation is related to the actual separation. Second, an
observer needs to assume a cosmological model to convert observed
angles and redshifts into comoving distances. The uniform
distribution of orientations is distorted if a wrong underlying
cosmology of the Universe is assumed.

\begin{figure}
\includegraphics[width=0.5\textwidth,height=0.25\textwidth]{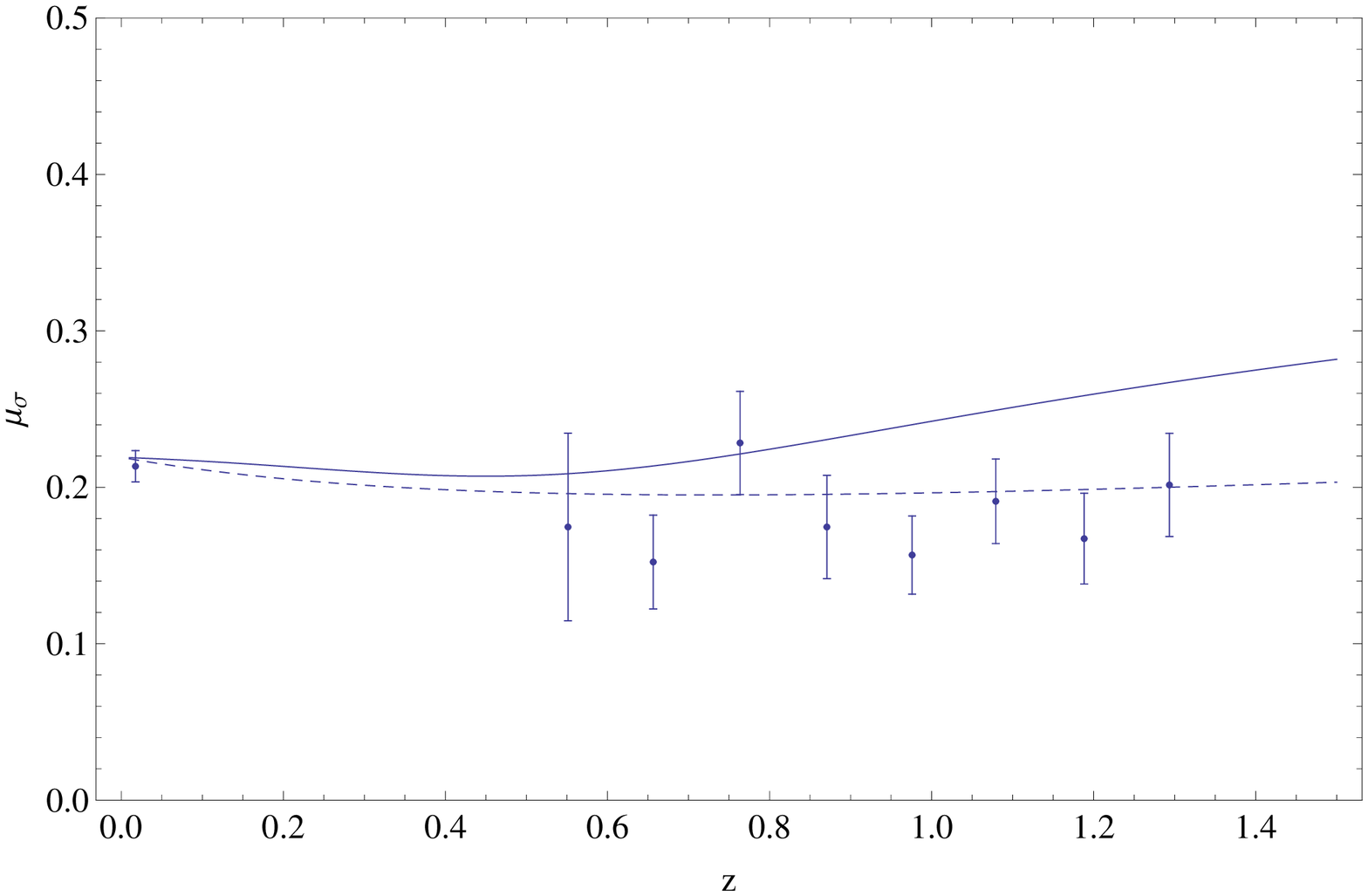}
\includegraphics[width=0.5\textwidth,height=0.25\textwidth]{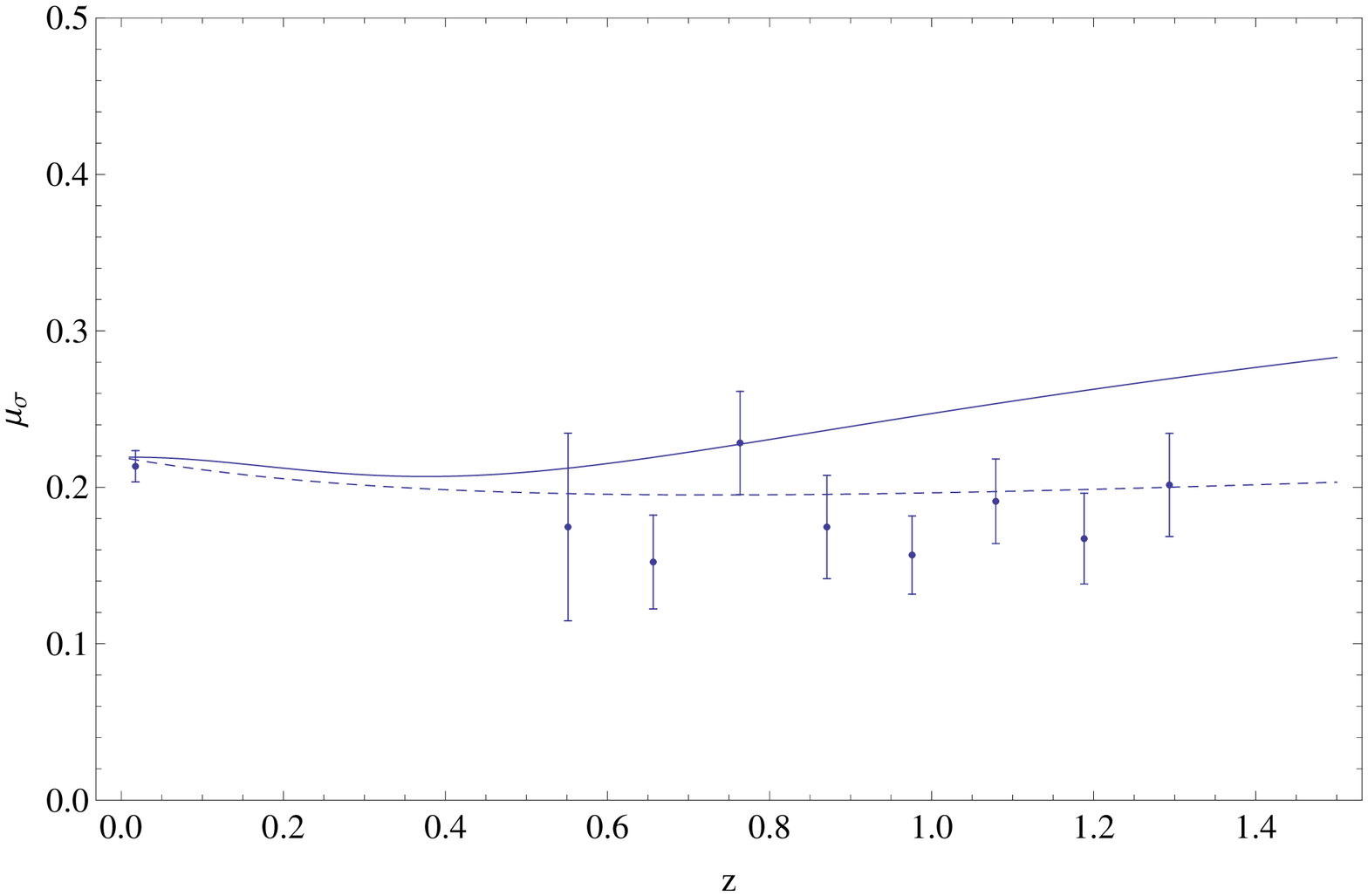}
\caption{\label{Fig3} Points represent the observed average
anisotropy of pairs. Solid lines represent the theoretical redshift
scaling of the AAP as predicted by Eq. (\ref{aapfunction}) in
different LTB models with best fit parameters from SNe Ia+CMB, up
panel for constrained GBH and bottom panel for gaussian LTB. The
dashed line shows the best fit $\Lambda$CDM model with
$\Omega_M=0.25$ and $\Omega_\Lambda$=0.65 }
\end{figure}

In a non-flat $\Lambda$CDM universe, the tilting angle $t$ subtended
between galaxy pairs and the line of sight, can be written as
\begin{equation}
\sin^2t=\{1+[C_k(\chi_A)\cot
\theta-\frac{S_k(\chi_A)C_k(\chi_B)}{S_k(\chi_B) \sin
\theta}]^2\}^{-1}, \label{tilt}
\end{equation}
for details, see Marinoni et al. (2012). It is nontrivial to
calculate the tilting angle and average anisotropy of pairs in LTB
models. The measured galaxy (matter) clustering and its evolution
agree with the standard $\Lambda$CDM cosmology to a factor of about
2 uncertainty up to z$\sim$1 (Tegmark et al. 2004; Coil et al. 2006;
Fu et al. 2008; Schrabback et al. 2010; Guzzo et al. 2008). A
minimalist approach is to simply use the $\Lambda$CDM value since
any viable LTB models must be consistent with these data. So we use
the observed average anisotropy of pairs from Marinoni \& Buzzi
(2010) derived in $\Lambda$CDM cosmology. Zhang \& Stebbins (2011)
also approximated the matter power spectrum by its form in a
standard $\Lambda$CDM cosmology. The observed tilting angle is
shifted to apparent angle $\tau$ because of the geometric
distortions induced by the peculiar velocities of the pair's
members. The probability distribution function of the apparent angle
$\tau$ which is  is given by Marinoni \& Buzzi (2010)
\begin{eqnarray}
\Psi(\tau){\rm{d}}\tau &=& \frac{1}{2}\frac{(1+\sigma^2)(1+ \tan^2
\tau)}{\left[ 1 + (1+\sigma^2)\tan^2 \tau \right]^{3/2}}|\tan \tau|
{\rm{d}} \tau \, , \label{dist}
\end{eqnarray}
and the parameter $\sigma$ depends on the cosmological expansion
history as
\begin{equation}
\sigma (z) = \alpha \frac{H_0(r)(1+z)}{H_\parallel(r,t)}.
\end{equation}
The normalization parameter $\alpha$ is given by $\alpha = H^{-1}_0
\left(\left \langle d v^2_{\parallel}/{d r^2} \right
\rangle\right)^{1/2}$. Because in the LTB metric, the transverse and
radial expansion rates are different, so the correct value must be
used in our calculations. When we use the galaxy pairs for AP test,
the velocity perturbation $\sigma$ used in equation (12) is related
to the peculiar motions of the pair members along the line of sight.
So we use radial expansion rate $H_\parallel$ to calculate the
velocity perturbation. Because the AP test is similar to BAO, we can
see this formula is also similar to the redshift interval $\delta z$
corresponding to the acoustic scale in the radial direction
(Garcia-Bellido \& Haugboelle 2008; Biswas et al. 2010; Marra \&
Paakkonen 2010). In the homogeneous $\Lambda$CDM model, Marinoni \&
Buzzi (2010) used the normal expansion rate $H(z)$. Marinoni \&
Buzzi (2010) derived the distribution $\Psi(\tau)$ as the average
anisotropy of pair (AAP), which is given by
\begin{eqnarray}
\mu_{\sigma} = \int \sin^2\tau\Psi(\tau){\rm{d}} \tau=
\frac{(1+\sigma^2)\,\mbox{arctan}(\sigma) - \sigma}{\sigma^3} \, .
\label{aapfunction}
\end{eqnarray}
At $z \approx 0$, Marinoni \& Buzzi (2010) obtained $\alpha =
5.79^{+0.32}_{-0.35}$, using binaries in the seventh data release of
the Sloan Digital Sky Survey (SDSS) (Abazajian et al. 2009). The
normalization factor $\alpha$ is assumed to be constant for all
redshifts and for different galaxy selections (Marinoni \& Buzzi
2010). Although Jennings et al. (2012) found that the value of
$\alpha$ could have a small variation with cosmology and redshift,
Marinoni \& Buzzi established that the changes of best fit value
cannot exceed the $1\sigma$ confidence level if the variation of
$\alpha$ is less than 10\%. So this assumption could be reasonable.
Belloso et al. (2012) also found that observations of close-pairs of
galaxies do show promise for AP cosmological measurements,
especially for low mass, isolated galaxies. The high-redshift (up to
$z\approx 1.45$) AAP are obtained using the third data release of
the DEEP2 survey (Davis et al. 2007). The value of $\chi'^2_{\rm
AAP}$ is
\begin{equation}
\chi'^2_{\rm AAP}=\sum_{i=1}^{9}
\frac{[\mu_{\sigma}-\mu_{\sigma,obs}(z_i)]^2}{\sigma_{obs,i}^2}.
\end{equation}
We adopt the value of $\mu_{\sigma,obs}$ and $\sigma_{\rm obs}$ from
Fig.~2 of Marinoni \& Buzzi (2010), which are shown as points in the
Fig~\ref{Fig3}. In order to verify the hypothesis that the
normalization factor $\alpha$ is constant for all redshifts, the
distance between the observed value of recession velocity difference
square $\langle dV^2_o(z) \rangle$ and the prediction of equation
(S20) in Marinoni \& Buzzi (2010) is minimal (see Marinoni \& Buzzi
(2010) for more details). So this $\chi^2$ value is
\begin{equation}
\chi'^2_{\rm dV^2}=\sum_{i=1}^{9} \frac{[\left \langle dV^2_o(z_i)
\right \rangle -\left \langle dV^2(z_i)\right
\rangle]^2}{\sigma_{dV_o^2,i}^2}.
\end{equation}
We use the value of  $\langle dV^2_o(z) \rangle$ and
$\sigma_{dV_o^2}$ from Fig.(5S) of Marinoni \& Buzzi (2010). The
total $\chi^2_{\rm AAP}$ is
\begin{equation}
\chi^2_{\rm AAP}=\chi'^2_{\rm AAP}+\chi'^2_{\rm dV^2}.
\end{equation}

In Fig.~\ref{Fig1}, we show the $1\sigma$, $2\sigma$ and $3\sigma$
contours in the $\Omega_{M,\rm in}-r_0$ plane for the constrained
GBH model. In the calculation, the priors from WMAP7, such as the
age of Universe $t_0=13.79$ Gyr and spectral index $n_s=0.96$ are
used (Komatsu, et. al. 2011). We also marginalize the Hubble
constant $H_0$ in the range $50\leq H_0 \leq 80~\rm
km~s^{-1}Mpc^{-1}$. The constraint from SNe Ia and CMB is shown as
dash-dot contours, and dashed contours for AAP. The allowed range of
$r_0$ is $1.80~{\rm Gpc}<r_0<4.10~\rm Gpc$ at $3\sigma$ level from
SNe Ia+CMB. But the allowed range of $r_0$ is $r_0>4.42$ Gpc at
$3\sigma$ level from AAP. These two contours do not overlap. So the
constrained GBH model can not explain the observations of SNe Ia+CMB
and AAP. The solid contours are derived from SNe Ia+CMB+AAP with
$\chi^2_{\rm min}=641.40$. While for the $\Lambda$CDM model, the
minimum $\chi^2$ is $620.37$. The constrained GBH model is excluded
at the $4\sigma$ confidence level compared to $\Lambda$CDM. In
Fig.\ref{Fig2}, we show the $1\sigma$, $2\sigma$ and $3\sigma$
contours in the $\Omega_{M,\rm in}-r_0$ plane for the gaussian LTB
model. The solid contours are derived from SNe Ia+CMB+AAP with
$\chi^2_{\rm min}=646.50$. This model is also excluded at the
$4\sigma$ confidence level compared to $\Lambda$CDM. From the
$\chi^2_{\rm min}$ of the two void models, we conclude that the
precise form of the density profile may not be essential. Because
the void models depend crucially on the void depth
$\delta_\Omega=(\Omega_{M,\mathrm{in}}-\Omega_{M,\mathrm{out}})/\Omega_{M,\mathrm{out}}$
and the void size $r_0$. So our conclusion is almost independent of
void model.

In Fig.\ref{Fig3}, we show the theoretical redshift scaling of the
AAP in these two LTB models. In the up panel, we use the best fit
parameters from SNe Ia+CMB for the constrained GBH model,
$r_0=3.0$~Gpc and $\Omega_{M,\rm in}=0.10$. Obviously, the predicted
values of AAP deviate from the observational values at high
redshift. The $\chi^2$ value is 37.35 for these nine data points. In
the bottom panel, $r_0=3.6$~Gpc and $\Omega_{M,\rm in}=0.12$ are
used for the Gaussian LTB model. The $\chi^2$ value is 41.96 for
these nine data points.

We must note that the local Hubble constant $H_{\rm loc}$ is also a
big obstacle to the void models. Because the measurement of the
Hubble constant is carried out mostly within a distance of roughly
$r_{\rm loc}\sim 200$ Mpc (Riess et al. 2011; Freedman et al. 2012),
we obtain the $H_{\rm loc}$ (Marra \& Paakkonen 2010)
\begin{equation}
H_{\rm loc}=\int_0^{r_{\rm loc}}H_0(r)4\pi r^2dr/(4\pi/3r_{\rm
loc}^3).
\end{equation}
In order to fit both the SNe Ia and CMB, the value of $H_{\rm loc}$
is $64 \pm 3.2~\rm km~s^{-1}Mpc^{-1}$ in the constrained GBH model
or $63\pm 3.5~\rm km~s^{-1}Mpc^{-1}$ in the Gaussian LTB model.
Riess et al. (2011) determined the Hubble constant with 3\%
uncertainty as $73.8\pm 2.4$ $\rm km~s^{-1}~Mpc^{-1}$. Freedman et
al. (2012) measured the Hubble constant as $74.3\pm 2.1$ $\rm
km~s^{-1}~Mpc^{-1}$.

\section{Discussions}
Previous investigations show that void models can fit a variety of
cosmological observations without containing dark energy because the
lack of homogeneity gives a great degree of flexibility. For
example, since the last scattering surface is far away from regions
where SNe Ia are observed, the property of inhomogeneity allows a
model to be constructed which provides different physical densities
in the regions from which these two sets of observational data are
drawn. So, the best way to constrain inhomogeneous models is using
several sets of data that measure a range of observables at
comparable redshifts. In this paper, we confront two general classes
of void models with observations of SNe Ia, CMB and orientations of
galaxy pairs. The redshifts of SNe Ia and orientations of galaxy
pairs are almost in the same range. We find that the these two void
profiles \textit{can not} fit both SNe Ia plus CMB data and
orientations of galaxy pairs simultaneously. We also show that the
two void models can fit both SNe Ia and CMB data, but at the expense
of a Hubble constant so low that they can also be ruled out. So our
results favor the Copernician principle. We must also note that our
results are obtained under some assumptions, such as the chosen
priors and void profile, which is also discussed in Biswas et al.
(2010). So the void models are ruled out at the $4\sigma$ confidence
level given the explored models and priors. But observations
challenge the void models (Biswas et al. 2010; Zibin \& Moss 2011).
Future galaxy surveys such as BigBOSS (Schlegel et al. 2011) will
provide improved precision of AAP function, placing much more strong
constraints on inhomogeneity.

\section*{ACKNOWLEDGMENTS}
We thank an anonymous referee for helpful comments and suggestions.
We have benefited from reading the publicly available code of Marra
\& Paakkonen (2010). This work is supported by the National Natural
Science Foundation of China (grants 11103007 and 11033002).

\end{document}